\documentclass[aps,floatfix,superscriptaddress,twocolumn,tightenlines,amsmath,amssymb,nofootinbib,10pt,longbibliography,prl]{revtex4-2}

\usepackage[utf8]{inputenc}
\usepackage[T1]{fontenc}
\usepackage{braket}
\usepackage{graphicx}
\usepackage{booktabs}
\usepackage{mathtools}	
\usepackage{amssymb}
\usepackage{amsmath}		
\usepackage{amsfonts}
\usepackage{mathrsfs}
\usepackage{color}
\usepackage{upgreek}
\usepackage{amsthm}
\usepackage{psfrag}
\usepackage{ifsym}
\usepackage{dsfont}
\usepackage{enumerate}
\usepackage{enumitem}
\usepackage{multirow} 
\usepackage{lipsum}
\usepackage{float}
\usepackage[separate-uncertainty=true]{siunitx}
\usepackage{setspace}
\usepackage[dvipsnames,svgnames]{xcolor}
\usepackage[normalem]{ulem}
\usepackage{changepage}
\usepackage{rotating}
\usepackage{hhline}

\usepackage{silence}
\usepackage{multirow}
\usepackage{booktabs}

\usepackage{hyperref}
\definecolor{refColour}{rgb}{0.4, 0.19, 0.28}
\definecolor{citeColour}{rgb}{0.5,0.5,0.0}
\definecolor{citeColour}{rgb}{0.41, 0.16, 0.38}
\definecolor{extraColour}{rgb}{0, 0, 0}
\hypersetup{colorlinks=true,linkcolor=blue,citecolor=blue}

\usepackage{titlesec}
\titleformat*{\subsection}{\normalfont\centering}
\titleformat*{\section}{\normalfont\centering\large}

\newcommand\blfootnote[1]{%
  \begingroup
  \renewcommand\thefootnote{}\footnote{#1}%
  \addtocounter{footnote}{-1}%
  \endgroup
}

\begin{document}

\title{\normalfont\LARGE{
{Conference key agreement in a quantum network}
}}

\author{Alexander Pickston$^{\dagger}$}
\affiliation{Institute of Photonics and Quantum Sciences, School of Engineering and Physical Sciences, Heriot-Watt University, Edinburgh EH14 4AS, United Kingdom}

\blfootnote{$^{\dagger}$ These two authors contributed equally. \\$^{*}$ Corresponding author: a.fedrizzi@hw.ac.uk}

\author{Joseph Ho$^{\dagger}$}
\affiliation{Institute of Photonics and Quantum Sciences, School of Engineering and Physical Sciences, Heriot-Watt University, Edinburgh EH14 4AS, United Kingdom}

\author{Andrés Ulibarrena}
\affiliation{Institute of Photonics and Quantum Sciences, School of Engineering and Physical Sciences, Heriot-Watt University, Edinburgh EH14 4AS, United Kingdom}

\author{Federico Grasselli}
\affiliation{Institut für Theoretische Physik III, Heinrich-Heine-Universität Düsseldorf, Universitätsstraße 1, D-40225 Düsseldorf, Germany}

\author{Massimiliano Proietti}
\affiliation{Institute of Photonics and Quantum Sciences, School of Engineering and Physical Sciences, Heriot-Watt University, Edinburgh EH14 4AS, United Kingdom}

\author{Christopher L. Morrison}
\affiliation{Institute of Photonics and Quantum Sciences, School of Engineering and Physical Sciences, Heriot-Watt University, Edinburgh EH14 4AS, United Kingdom}

\author{Peter Barrow}
\affiliation{Institute of Photonics and Quantum Sciences, School of Engineering and Physical Sciences, Heriot-Watt University, Edinburgh EH14 4AS, United Kingdom}

\author{Francesco Graffitti}
\affiliation{Institute of Photonics and Quantum Sciences, School of Engineering and Physical Sciences, Heriot-Watt University, Edinburgh EH14 4AS, United Kingdom}

\author{Alessandro Fedrizzi}
\affiliation{Institute of Photonics and Quantum Sciences, School of Engineering and Physical Sciences, Heriot-Watt University, Edinburgh EH14 4AS, United Kingdom}

    \begin{abstract}
        Quantum conference key agreement (QCKA) allows multiple users to establish a secure key from a shared multi-partite entangled state.
        In a quantum network, this protocol can be efficiently implemented using a single copy of a N-qubit Greenberger-Horne-Zeilinger (GHZ) state to distil a secure N-user conference key bit, whereas up to N-1 entanglement pairs are consumed in the traditional pair-wise protocol.
        We demonstrate the advantage provided by GHZ states in a testbed consisting of a photonic six-user quantum network, where four users can distil either a GHZ state or the required number of Bell pairs for QCKA using network routing techniques.
        {In the asymptotic limit, we report a more than two-fold enhancement of the conference key rate when comparing the two protocols. We extrapolate our data set to show that the resource advantage for the GHZ protocol persists when taking into account finite-key effects.}
    \end{abstract}
    
\maketitle

    One of the great promises of quantum technology is the development of quantum networks, which will allow global distribution of entanglement for tasks such as distributed quantum computing~\cite{beals_efficient_2013, van_meter_path_2016}, distributed quantum sensing~\cite{gottesman2012, komar_quantum_2014} and quantum-secure communication~\cite{bennett_quantum_2014, ekert_quantum_1991, peev2009secoqc, sasaki2011tokyo, dynes2019cambridge, chen_implementation_2021}.
    To leverage the full potential of quantum networks we require protocols that draw an efficiency advantage from genuine multi-partite entanglement as opposed to strictly pair-wise correlations such as Bell states.
    Multi-user entanglement such as Greenberger-Horne-Zeilinger (GHZ) states have already found application in quantum conference key agreement~\cite{epping_multi-partite_2017, grasselli_finite-key_2018, murta_quantum_2020, proietti_experimental_2021}, quantum secret sharing~\cite{hillery1999, tittel2001, broadbent2009} and quantum communication complexity problems~\cite{buhrman2010, ho2022}.
    
\begin{figure}[ht!]
        \centering
        \includegraphics{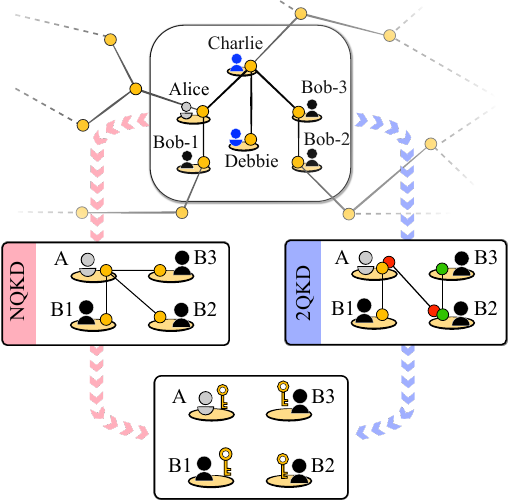}
        \caption{\textbf{Conceptual illustration of a multi-node quantum network.}
        Physical qubits are represented by nodes (circles), while edges (lines) represent pair-wise interactions that entangle the qubits within the wider network.
        Here, we define a {network} consisting of six users represented by nodes within the rounded rectangle.
        Within this {network}, four of the users may wish to perform quantum protocols such as conference key agreement.
        The traditional method involves distributing Bell pairs among the participants first, followed by post-processing steps to extract keys in a pairwise fashion before arriving at the final conference key as shown by the right path in blue.
        An alternative approach is to obtain a GHZ state and employ a multi-party quantum conference key agreement protocol to extract the key directly, as represented in the left red path.
        }
        \label{fig:concept}
    \end{figure}

    In quantum conference key agreement (QCKA), N users aim to establish a common and identical secret key for group-wide encryption. 
    Using standard two-party QKD schemes, this can be achieved by generating a set of N-1 pair-wise keys either via prepare-and-send schemes or via shared entangled Bell pairs~\cite{wengerowsky_entanglement-based_2018, joshi_trusted_2020}. The set of keys can then be used to algorithmically generate the conference key by performing a bit-wise XOR operation on the individual keys---we will refer to this as the `2QKD' approach.
    If the participating users are part of a network as depicted in Figure~\ref{fig:concept}, a more resource-efficient protocol called `NQKD' leverages multi-partite entanglement to obtain conference key bits directly from GHZ states supplied to all participating users.
    NQKD can outperform the 2QKD scheme by consuming up to N$-1$ times fewer network resources in constrained quantum networks~\cite{epping_multi-partite_2017,hahn_limitations_2022-1}.
    QCKA was recently demonstrated in a four-user NQKD scenario where a four-photon GHZ state was transmitted over up to 50~km of telecom fibre~\cite{proietti_experimental_2021}.
    However, a direct comparison showing the experimental resource advantage over the 2QKD approach in a quantum network has not yet been achieved.
    
    In this work we consider the scenario of multi-user conference key agreement in a small-scale quantum network, as illustrated in Figure~\ref{fig:concept}.
    The network comprises six users: Alice and three Bobs distil a secure conference key, while two non-participants, Charlie and Debbie, facilitate the operation of the network.
    This scenario, wherein participants of a protocol are indirectly connected via other non-participating users, is expected to naturally occur in future multi-node quantum networks that generate large-scale entanglement in the background.
    In order to efficiently distribute entanglement resources to requesting parties, we will use so-called network coding techniques~\cite{hein_multiparty_2004,hahn_quantum_2019,adcock_mapping_2020} which involve local operations, quantum measurements and classical communication to alter the network connectivity.
    Using this network we implement the two QCKA approaches, i.e., NQKD and 2QKD, to demonstrate the resource advantage when using the multi-partite-entanglement-enabled protocol.
    
    For the NQKD approach we consider the N-BB84 protocol~\cite{grasselli_finite-key_2018} that was implemented in Ref.~\cite{proietti_experimental_2021}.
    In each protocol round, an N-partite GHZ state is distributed among N users.
    Each user performs a measurement on their respective qubit according to a pre-agreed sequence of two possible measurements.
    In type-1 rounds each user measures their qubit in the Pauli $\text{Z}$-basis, exploiting the perfect Z-correlations for generation of the raw key; type-2 rounds correspond to jointly measuring in the Pauli $\text{X}$-basis and are used for parameter estimation.
    Security is established by evaluating the phase error rate ($Q_\text{X}$) from the type-2 rounds and determining the quantum bit error rate (QBER) from a random subset of publicly disclosed type-1 outcomes.
    We can assess the performance of the N-BB84 protocol, in the limit of an infinite number of rounds, by determining these two parameters and calculating the asymptotic key rate (AKR)~\cite{grasselli_finite-key_2018, proietti_experimental_2021}. 
    The AKR establishes the fractional secret key bit extracted for each copy of the resource state and is defined as,
	\begin{equation}
	    \mathrm{AKR}=1-H(\textrm{QBER})-H(Q_{\text{X}}),
	    \label{eqn:akr}
	\end{equation}
	where $H(x)\doteq-x\log_2 (x) -(1-x)\log_2 (1-x)$ is the binary entropy function.
	The explicit evaluation of QBER and $Q_{\text{X}}$ is detailed in the Methods.
	
	In the case of 2QKD, users initially obtain sets of Bell states, rather than a GHZ state, from which they run pairwise BB84 protocols whose AKR is also given by the expression in~(\ref{eqn:akr}).
	However, unlike in the NQKD approach, the 2QKD method involves obtaining N$-1$ unique, pairwise keys spanning the N-user group first. 
	This is followed by a classical step, e.g., applying the bit-wise XOR, to transform the individual keys into the final conference key~\cite{epping_multi-partite_2017}.
	As we will show in our experiment, depending on the underlying topology of the quantum network, this leads to the 2QKD approach requiring up to N$-1$ times more network resource states than NQKD.

    \begin{figure*}[ht!]
        \centering
        \includegraphics{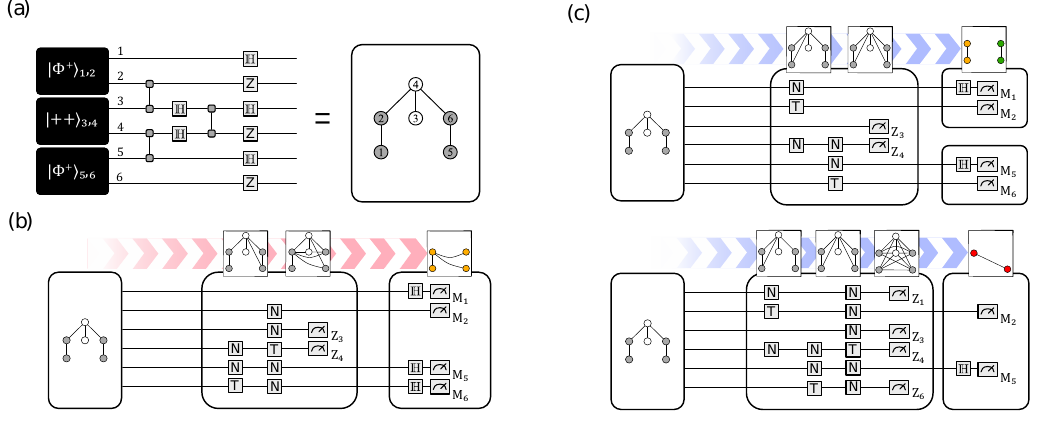}
        \caption{\textbf{Experimentally generating and manipulating the six-photon graph state.} 
        (a) On the left is the linear optical circuit for generating the target graph state on the right.
        Three photon-pair sources are represented as black boxes with each horizontal line representing a labelled optical mode containing one photon.
        Two sources produce Bell states, $|\Phi^+\rangle$, while the third generates a biseparable state with each photon in $|+\rangle$.
        The two-qubit fusion gates are denoted by two squares on the modes they act upon connected by a vertical line.
        Single-qubit operations, i.e., Hadamard and Pauli-Z gates are shown as squares with letters $\mathds{H}$ and Z respectively.
        (b) The set of graph transformations for obtaining the four-qubit GHZ state, in modes $\{1,2,5,6\}$, are depicted in the red path.
        This corresponds to local operations consisting of single-qubit gates $\textrm{N} \doteq \sqrt{-i\text{X}}$ and $\textrm{T} \doteq \sqrt{i\text{Z}}$, where $\text{X}$ and $\text{Z}$ are again Pauli gates, followed by quantum measurements on non-participatory qubits $\{3,4\}$ in the $\text{Z}$ basis.
        The remaining qubits can be measured in the joint-$\text{Z}$ or-$\text{X}$ basis allowing us to evaluate the key rate performance for the NQKD method.
        (c) Local-graph operations for obtaining Bell pairs between nodes $\{1,2\}$ and $\{5,6\}$ (top) and a Bell pair between nodes $\{2,5\}$ (bottom) which are needed for the 2QKD approach.
        }
        \label{fig:expperimentalTechnique}
    \end{figure*}

\vspace{1em}
\noindent
\textbf{Results}
\newline
    Quantum networking studies usually assume a basic universal building block, for example the `ring' network from which larger networks can be constructed.
    A common feature of these building blocks is that they allow arbitrary pairs of nodes to share Bell pairs via network coding routines, which distill sub-graphs from a network via local measurements and the discarding of non-participating network nodes.
    The six-node quantum network we implement, Figure~\ref{fig:expperimentalTechnique}~(a), shares several features with the universal ring network, allowing us to directly compare GHZ-based QCKA and the Bell-pair-based 2QKD. 
    We experimentally constructed the 6-photon network via a linear optics setup consisting of telecom-wavelength photon-pair sources and polarisation optics, see Methods for details.
    
    We now outline the method of manipulating the 6-node graph state~\cite{hein_multiparty_2004} via local complementation (LC)~\cite{hein_multiparty_2004, adcock_mapping_2020, hahn_quantum_2019}.
    The LC operations on the six-photon graph results in a four-party GHZ state in modes $\{1, 2, 5, 6\}$, corresponding to four users, e.g., Alice, Bob-1, Bob-2 and Bob-3, implementing NQKD is shown in Figure~\ref{fig:expperimentalTechnique}(b).
    Conventionally, Alice denotes the party who co-ordinates the error correction and privacy amplification steps after the raw key is established~\cite{proietti_experimental_2021, grasselli_finite-key_2018, epping_multi-partite_2017}.
    We also include LC operations mapped to sets of single-qubit gates~\cite{adcock_mapping_2020} in the circuit.
    Nodes 3 and 4 represent users Charlie and Debbie, who do not participate in the QCKA key generation.
    They measure their qubits in the $\text{Z}$ basis then announce their outcomes, which allows the qubits to be coherently removed from the graph.
    The remaining four qubits are transformed into the star graph which is a locally equivalent to the four-party GHZ state~\cite{hein_multiparty_2004} required for the NQKD protocol.
    We then construct measurement sequences consistent with type-1 and type-2 rounds, i.e., $\text{Z}^{\otimes4}$ and $\text{X}^{\otimes4}$ respectively, of the NQKD protocol.
    
    The method for allocating Bell pairs in the network, for the 2QKD protocol, is shown in Figure~\ref{fig:expperimentalTechnique}(c).
    To generate a conference key among the four users, at least three pairwise keys are required.
    There are two sets of network transformations for obtaining three individual Bell pairs among the same group of four users that participated in the NQKD protocol.
    The top of Figure~\ref{fig:expperimentalTechnique}(c) shows how to obtain Bell pairs between nodes $\{1,2\}$ and $\{5,6\}$ respectively from a single copy of the {network resource state.}
    The bottom panel shows how a single Bell pair is established between modes $\{2,5\}$. 
    We similarly construct the measurement sequences corresponding to type-1 and type-2 rounds for each Bell pair, see Methods for details.

    \begin{table}[b!]
    \setlength{\tabcolsep}{8pt} 
    \renewcommand{\arraystretch}{1.25} 
    {\small
    \begin{center}
        \begin{tabular}{c|c|c|c}
        State & QBER & Q$_{X}$ & AKR \\ \hline
        GHZ & 0.080(5) & 0.11(11) & 0.093(22) \\
        Bell$_{1,2}$ & 0.076(5) & 0.079(13) & 0.22(3) \\
        Bell$_{5,6}$ & 0.074(5) & 0.086(13) & 0.20(3) \\
        Bell$_{2,5}$ & 0.102(5) & 0.100(13) & 0.057(3)
        \end{tabular}
    \end{center}
    \caption{{\textbf{Noise terms measured for each of the states in the QCKA protocol.} Here, we report the noise parameters as well as the resulting asymptotic key rates for each of the states derived from the network.}}
    \label{tab:results}
    }
    \end{table}
    
    The outcomes of type-1 and type-2 rounds are used to calculate the noise parameters QBER and $Q_\text{X}$ which are then used to evaluate the AKR using the expression in (\ref{eqn:akr}).
    This expression corresponds to the asymptotic conference key rate for the NQKD approach and we denote it as $\textrm{AKR}_\text{N}$.
    For the 2QKD approach, we use Eq.~(\ref{eqn:akr}) to compute the pairwise AKR of each BB84 protocol, denoted $\{r_{AB_1}, r_{B_2B_3}, r_{AB_2}\}$; the resulting asymptotic conference key rate is then obtained using the expression,
    \begin{equation}
        \textrm{AKR}_{2} = \frac{1}{\frac{1}{r_{AB_2}} + \textrm{max} \{ \frac{1}{r_{AB_1}}, \frac{1}{r_{B_2B_3}} \}} .
        \label{eqn:akr2qkd}
    \end{equation}
    In the ideal case where the AKR of each Bell pair is unity, we obtain $\textrm{AKR}_{2}^{ideal} = 1/2$ while for the NQKD case with an ideal four-party GHZ state we can attain $\textrm{AKR}_{\text{N}}^{ideal} = 1$.
    The expected ratio of the key rate advantage in favour of NQKD is therefore $\textrm{AKR}_{\text{N}}^{ideal}:\textrm{AKR}_{2}^{ideal} = 2$.
    This advantage originates from the ability to use a single copy of the network resource state to produce one secure bit of the conference key via NQKD, whereas in 2QKD each secure bit requires two copies of the network resource state.

    \begin{figure*}[ht!]
        \centering
        \includegraphics{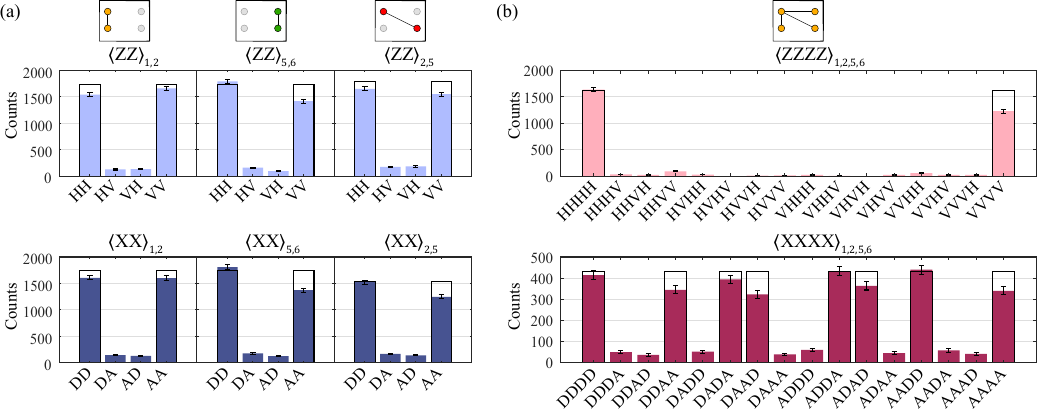}
        \caption{\textbf{Experimentally measured populations of the different resource states for NQKD.} 
        (a) Populations of the three Bell pairs used for 2QKD.
        (b) Populations of the four-party GHZ state for NQKD.
        The type-1 rounds are shown in the top set of measurements, while type-2 rounds are shown in the bottom set.
        These outcomes are extracted from complete measurements in the six-photon state space, wherein post-selection of the non-participatory nodes has taken place.
        The theoretically expected populations are included as transparent bars.
        All measurements are recorded for the same integration time {(300 seconds)} at a fixed optical pump power {(75 mW)} and error bars are shown for one standard deviation assuming Poissonian statistics.
        {We collected $4801$ rounds of data, which in total took $\sim400$ hours, accumulating $20272$ six-fold counts} 
        }
        \label{fig:results}
    \end{figure*}

    We compare the experimental results of type-1 and type-2 measurements for both approaches in Figure~\ref{fig:results}.
    We set the pump power to each source to obtain an average six-photon detection rate of $\sim 0.0141$~Hz for all measurements.
    This optical pumping regime was chosen to limit multi-photon events while still ensuring sufficient detection rates to collect statistics for each protocol, for details on the effect of optical pump parameters see Supplementary Materials.
    From the data in Figure~\ref{fig:results} we calculate QBER and $Q_\text{X}$, for each resource state.
    We then evaluate the asymptotic conference key rate for NQKD, $\textrm{AKR}_{\text{N}}=0.093(22)$ and for 2QKD, $\textrm{AKR}_{2}=0.044(15)$.
    We arrive at the experimentally measured ratio of the two approaches to be $\textrm{AKR}_{\text{N}}^{exp}:\textrm{AKR}_{2}^{exp}=2.13(6)$.
    The uncertainties reported here are taken as one standard deviation from the average of samples via a Monte Carlo simulation which assumes Poissonian counting statistics.
    
\begin{figure}[h!]
        \centering
		\vspace{-1em}
        \includegraphics{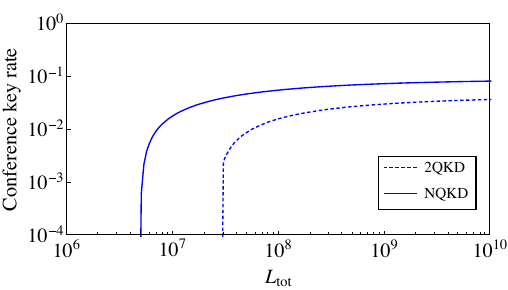}
        \vspace{-2em}
        \caption{\textbf{Conference key rate, with finite-key effects, as a function of the number of successful network usages ($L_{tot}$).}
        The performance for the NQKD protocol is shown by the solid line, while the 2QKD protocol is given by the dashed line. Both assume a fixed total security parameter, $\epsilon_{tot}=10^{-8}$.
        \label{fig:FKR}}
    \end{figure}

    In practice, only a limited number of rounds can be performed and this necessitates accounting for finite-key effects in the conference key rate~\cite{proietti_experimental_2021}.
    We perform a finite-key analysis for both NQKD and 2QKD by using the measured noise parameters for each resource state, see Table~\ref{tab:results}, and by simulating the finite-key effects for a range of total rounds, $L_{tot}$, where each round corresponds to a successful distribution of the shared resource state.
    The finite key rate expression for the NQKD approach is explicitly defined in Ref.~\cite{proietti_experimental_2021,grasselli_finite-key_2018}, while its calculation for 2QKD presents some notable differences. In the 2QKD approach, the first step is to obtain three separate pair-wise keys, whose length is given by the finite-key rate formula of the BB84 protocol (it can be recovered from the NQKD finite-key rate by setting the number of parties to two). The second step employs the established keys to distribute the final conference key via a one-time pad.
    As a result, the conference key length cannot exceed the shortest pair-wise key among the three established ones. Moreover, for a given security parameter associated to the conference key, $\epsilon_{tot}$, each bipartite key must be processed with a more stringent security parameter in order to recover $\epsilon_{tot}$ through composability, which reduces the overall key rate. Furthermore, in 2QKD the total number of network resource states is subdivided among the three pairwise protocols, implying that each protocol can only rely on a smaller set of data, thereby increasing the weight of the statistical corrections in the key rate. By these arguments, we expect the NQKD approach to retain its advantage over 2QKD in the finite-key regime. Figure~\ref{fig:FKR} reports the conference key rates with finite-key effects of both NQKD and 2QKD, where the total security parameter is fixed to: $\epsilon_{tot}=10^{-8}$.
    We observe that the advantage of the NQKD protocol over the 2QKD counterpart increases significantly in the finite-key regime. Remarkably, the minimal number of resource states required to distil a non-zero conference key is reduced by nearly one order of magnitude with NQKD.
    
\noindent
\textbf{Discussion}
\newline
    We have shown that, once genuine quantum networks capable of providing multi-node entanglement in the background become available, NQKD conclusively outperforms 2QKD in terms of key rate per network use.
    The NQKD advantage is at best $\text{N}-1$. Many graphs however will allow for a Bell state \emph{multi-cast}~\cite{epping_multi-partite_2017, hahn_quantum_2019}, where more than one Bell pair can be distilled in a single network use between participating users. The graph we chose for our network demonstration allows precisely this, reducing the maximal theoretical NQKD advantage over 2QKD from three-fold to two-fold, thus providing the fairest possible comparison between the two QCKA protocols.
    In the Supplementary Discussion we investigate the option to multi-cast in the ring network.

    Conversely, in a scenario with \emph{direct transmission} of GHZ states over multiple quantum channels connecting Alice to the Bobs (as in ~\cite{proietti_experimental_2021}), NQKD is unlikely to achieve higher absolute rates than what is possible via 2QKD with state-of-the-art systems.
    GHZ state generation is currently probabilistic and slow, but will eventually catch up because deterministic multi-photon sources are on the development path for optical quantum computation.
    Even then, NQKD requires N photons to be detected simultaneously, which exponentiates the known rate-distance limit~\cite{pirandola_direct_2009, pirandola_fundamental_2017} in individual quantum channels.
    In 2QKD, N$-1$ photons can be transmitted and detected asynchronously, and the rate loss is therefore merely additive.
    This comparison becomes more complex once repeater nodes are incorporated into networks, as multi-hops and routing configurations need to be considered~\cite{pirandola_end--end_2019-1, pirandola_general_2020, ottaviani_modular_2019-1}.
    Indeed, when allowing for storage of quantum states at repeater nodes the transmission of GHZ states instead of Bell pairs can improve transmission distances, storage times and robustness to noise~\cite{miguel-ramiro_optimized_2023,wallnofer_multipartite_2019,kuzmin_scalable_2019}.
    
    As one of the first multi-user quantum communication protocols with a network advantage, QCKA is now a very active research area.
    Protocols have been developed for measurement-device independent scenarios~\cite{ottaviani_modular_2019-1, Fu_2015} and implementations with weak coherent states that might outperform N-photon GHZ state transmission~\cite{grasselli_wStateCKA_2019, Cao_2021}.
    Loss-resilient encoding of GHZ states for QCKA using error correction methods have also been proposed~\cite{singkanipa2021}.
    Studies of QCKA have raised fundamental questions about the type of entanglement that is useful in multi-user quantum communication.
    It has been shown that genuine multi-partite entanglement is not strictly required for QCKA~\cite{carrara_genuine_2021}.
    Nonetheless, obtaining a non-zero conference key could be used as a witness for entanglement in a network~\cite{carrara_genuine_2021, Das_2021}.
    
    It will be important to explore other network protocols that gain an advantage from multi-partite entanglement. 
    One example is QCKA-based \emph{anonymous} conference key agreement which generates a conference key while hiding the identities of the participants~\cite{hahn2020,grasselli_secure_2022,de2022anonymous,thalacker2021anonymous}, theoretically achieving a significant advantage when using both GHZ states and Bell pairs~\cite{grasselli_secure_2022} compared to using Bell pairs alone~\cite{huang2022_anonymous}.
    Quantum secret sharing~\cite{walk2021} also gains a network advantage from GHZ states.
    Going beyond communication protocols, it has been shown that distributed sensing can benefit from multi-partite entanglement shared between measurement nodes~\cite{liu_distributed_2021}.
    A key question for these protocols is whether the required multi-party sub-graphs can be obtained from a network efficiently.
    For networks based on graph states this has been answered in the affirmative~\cite{hahn_quantum_2019} however generalisations to other resource states is an ongoing challenge~\cite{dahlberg_how_2020}.
\vspace{1em}
\noindent
\textbf{Acknowledgements}
\newline
    This work was supported by the UK Engineering and Physical Sciences Research Council (Grant Nos. EP/T001011/1.). 
    F Graffitti acknowledges studentship funding from EPSRC under Grant No. EP/L015110/1. 
    F Graselli acknowledges support from the Deutsche Forschungsgemeinschaft (DFG, German Research Foundation) under Germany’s Excellence Strategy - Cluster of Excellence Matter and Light for Quantum Computing (ML4Q) EXC 2004/1 -390534769.

\vspace{1em}
\noindent
\textbf{Author contributions}
\newline
    AP and JH are co-first authors.
    MP, AF conceived the project. 
    F Grasselli, AP, and JH developed the theoretical framework. 
    AP, JH, AU, CLM, F Graffitti, and PB performed the experiment and collected the data. 
    AP, JH analysed experimental data and prepared figures. 
    All authors contributed to writing the manuscript.

\vspace{1em}
\noindent
\textbf{Competing interests}
\newline
    The authors declare no competing interests.

\clearpage
\newpage

{\small
\noindent
{\normalsize\textbf{Methods}}
\newline
\noindent
\textbf{Parameter estimation.}\label{subsec:paramEst} 
    In the N-BB84 protocol QBER is defined as the largest pair-wise error rate observed from the type-1 rounds, i.e., when all users measure jointly in the $\text{Z}$ basis.
    For convenience we adopt a general labelling of a group of $\text{N}$ users following the set, $\{A,B_1, B_2,...,B_{\text{N}-1}\}$.
    Thus for an $\text{N}$-GHZ state, errors in the $\text{Z}$ basis occur when there is a non-zero probability that one of the Bobs obtains an outcome that differs from Alice. Explicitly, the QBER is defined as:
	\begin{equation}
	    \textrm{QBER} \doteq \max_{i\in\{1,2,\dots,(\text{N}-1)\}} Q_{AB_i} ,
	\end{equation}
	where $Q_{AB_i}=\Pr(\text{Z}_A \neq \text{Z}_{B_i})=(1 - \langle \text{Z}^A \text{Z}^{B_i} \rangle)/2$.
	When evaluating the pair-wise error rates for different $i$, the role of Alice does not have to be assigned to a specific user.
	Rather, we can permute the role of Alice within the set of users, by updating the indices of each user.
	We find different maximum pair-wise error rates (QBER) depending on the permutation, due to noise being distributed unevenly among each of the six qubits of the resource state.
	By iterating through all permutations, we select the role of Alice based on yielding the lowest QBER, optimising the conference key rate.
	
	Phase error for a $\text{N}$-GHZ state can be estimated from $\text{X}$ basis measurements. 
	This is calculated by determining the deviation in measurement outcome from the expected correlations.
	This is expressed as, 
	\begin{equation}
		Q_{\text{X}} = (1 - \braket{\text{X}^{\otimes \text{N}}})/2.
	\end{equation}
	Both QBER and $Q_{\text{X}}$ are zero in the absence of error and noise, indicating that there is no deviation from the expected behaviour of the GHZ state.
	
	For Bell states, both QBER and $Q_\text{X}$ can be evaluated using the same expressions as above.
	However, since there are only two users sharing a Bell state we skip the step of permuting the role of Alice in the group.

\vspace{1em}
\noindent
\textbf{Experimental layout.}\label{subsec:ExpSet}\emph{ Photon pair sources---}
    To produce the six-photon graph state we use three photon-pair sources which are optically pumped by a {$1.3$~}picosecond Ti-Sapph laser {that has a nominal repetition rate of $80$~MHz}.
    All three are type-II parametric down-conversion (PDC) sources based on domain-engineered aperiodically-poled KTP (aKTP) crystals that are optimally phase-matched for spectrally pure photon pairs at 1550~nm.
    {We can therefore achieve two-photon interference visibilities of up to $98.6\pm1.1\%$.}
    {For more details on the design of the crystals, please see Ref.~\cite{pickston_optimised_2021}.}
    High spectral purity is achieved without needing narrow-band filters which improves collection efficiency and increases the overall detection rate. 
    Each aKTP crystal is embedded in a polarisation-based Sagnac interferometer which can be optically pumped bidirectionally to produce polarisation-entangled photon pairs~\cite{fedrizzi_wavelength-tunable_2007},
    \begin{equation*}
        | \Psi^- \rangle_{a,b} = \frac{|\textsf{hv}\rangle_{a,b} - |\textsf{vh}\rangle_{a,b}}{\sqrt{2}},
    \end{equation*}
    where $|{\textsf{h}}\rangle$ and $|{\textsf{v}}\rangle$ correspond to horizontal and vertical polarisation states respectively, and subscripts denote optical modes.
    We remark that the $|\Psi^-\rangle$ Bell state can be transformed into the $|\Phi^+\rangle$ Bell state via a local operation using polarisation optics.
    {Both Sagnac interferometers produced Bell states with purities of $0.9783(5)$ and $0.9706(3)$.} 
    Lastly, we can configure one of the sources to produce separable states by optically pumping the Sagnac loop in one direction, thus generating the state $|++\rangle \doteq |\textsf{dd}\rangle$, where $|\textsf{d}\rangle \doteq \left( |\textsf{h}\rangle + |\textsf{v}\rangle\right) /\sqrt{2}$ is the diagonal polarisation state.
    
    The linear optical circuit responsible for producing the six-photon graph state consists of three fusion gates~\cite{browne_resource-efficient_2005} along with single-qubit gates on the respective modes, shown in Figure~\ref{fig:expperimentalTechnique}.
    The linear optical fusion gate is probabilistic, with a success rate of $1/2$, however successful operation is heralded by the detection of one photon in each output of the gate.
    {The overall success probability of all three fusion gates is $1/8$.}
    See Supplementary Methods for further experimental detail.
    After the network of fusion gates and upon detecting one photon in each of the six numbered modes, the joint quantum state of the system is given by,
	\begin{equation}
	\begin{aligned}
		\ket{\Psi}_{\text{initial}} = &(\ket{\textsf{hhhhhh}} - \ket{\textsf{hhhhvv}} - \ket{\textsf{hhvvhh}} \\
		& - \ket{\textsf{hhvvvv}} -  \ket{\textsf{vvhhhh}} + \ket{\textsf{vvhhvv}} \\
		& - \ket{\textsf{vvvvhh}} - \ket{\textsf{vvvvvv}})/\sqrt{8}.
	\end{aligned}
	\end{equation}
    We remark that this state is equivalent to the six-photon graph state discussed in the main body of work, subject to a set of local rotations,
    $$(\mathds{H} \otimes \text{Z} \otimes \mathds{H} \otimes \text{Z} \otimes \mathds{H} \otimes \text{Z}) \ket{\Psi}_{\text{initial}} = \ket{\textsf{G}}_{\text{target}}$$
    where the target state is a graph state,  $\ket{\textsf{G}}_{\text{target}}$ and $\mathds{H}$ is the Hadamard gate.

    After the fusion gates, the photons go to tomography stages which consist of a HWP, QWP and a PBS, after which photons are coupled into single mode fibres.

\vspace{1em}
\noindent    
\textbf{Graph state formalism.}
    We use the graph state formalism to represent the transformations of the resource state used to distribute the GHZ state and Bell pairs between the four users.
    As the resource state is capable of distilling a GHZ state and sets of Bell pairs, this enables direct comparison of QCKA based on NQKD and 2QKD without changing the structure of the linear optical circuit.
	The general form of a graph state $\textsf{G} = (V,E)$ reads:    
	\begin{equation}\label{eq:generalGraph}
		\ket{\textsf{G}} = \prod_{(i,j) \in E} \textsf{CZ}_{ij} \ket{+}^{\otimes |V|}.
	\end{equation}
	where $\textsf{CZ}_{i,j}$ is the two-qubit controlled-Z gate acting on modes $\{i,j\}$, $E$ is the list of edges connecting two vertices and $|V|$ is the number of vertices of the graph.
	Using this notation, we define our target graph state as,
    \begin{equation}
        \begin{aligned}
            \ket{\textsf{G}}_{\text{target}} = (\textsf{CZ}_{1,2} \otimes \textsf{CZ}_{2,4} \otimes \textsf{CZ}_{3,4} \otimes \textsf{CZ}_{4,6} \otimes \textsf{CZ}_{5,6}) \ket{+}^{\otimes 6},
        \end{aligned}
    \end{equation}
    For more details on this construction and the graph state formalism see Supplementary Methods.
    
	One specific transformation within the graph state formalism is called local complementation (LC). 
    An LC operation itself is comprised of two different unitaries. 
    On a vertex designated as the target, the operation $\text{T}=\sqrt{-i\: \text{X}_t}$ is applied {where, $$\text{T}=\left(\begin{smallmatrix} \frac{1}{\sqrt{2}} & -\frac{i}{\sqrt{2}} \\ -\frac{i}{\sqrt{2}} & \frac{1}{\sqrt{2}}\end{smallmatrix}\right)$$} whilst each vertex neighbouring the target, the operation $\text{N}=\sqrt{i\:\text{Z}_n}$ is applied {where, $$\text{N}=\left(\begin{smallmatrix} e^{\frac{i \pi }{4}} & 0 \\ 0 & -i e^{\frac{i \pi }{4}}\end{smallmatrix}\right).$$}
    
\vspace{1em}
\textit{Measurement sequences---}\label{subsec:measSeq}
    After the fusion gates, each photon is sent to a tomography stage consisting of a set of controllable HWP and QWP then onto a PBS with both output ports fibre coupled.
    From the initial target graph $\ket{\textsf{G}}_{\text{target}}$ LC operations are used, transforming the initial graph into alternative graphs within the same entanglement class (or graph state orbit), whereby different combinations of Bell states or a single GHZ state can be obtained~\cite{adcock_mapping_2020}.
    This is illustrated in Figure~\ref{fig:expperimentalTechnique} in the main text.
    
    We assign Alice, Bob-1, Charlie, Debbie, Bob-2 and Bob-3, to qubit numbers 1, 2, 3, 4, 5 and 6 respectively.
    To evaluate the AKR for a comparison between QCKA techniques, we need to calculate $Q_{\text{Z}}$ (or QBER) and $Q_{\text{X}}$ for each state that plays a part in the QCKA protocol.
    All single-qubit operations applied to the initial graph such that network users  obtain the correct state and perform the correct measurements are encoded onto the measurement settings, including the LC operations.
    Obtaining a GHZ shared between Alice and the Bobs in the four-photon state space, to carry out QCKA with a multi-partite resource, Charlie and Debbie are required to make projections in the $\text{Z}$ basis. 
    So for each round of measurements, whilst Charlie and Debbie should always measure in the $\text{Z}$ basis, Alice and the Bobs measurements are made based on whether they want to complete a type-1 round or type-2 round of the protocol.
	For example, to measure the observable $\langle \text{Z}_1 \text{Z}_2 \text{Z}_5 \text{Z}_6 \rangle$ of the GHZ state to evaluate the QBER, all users must measure the observable $\langle \text{Z}_1 \text{Z}_2 \text{X}_3 \text{X}_4 \text{Z}_5 \text{Z}_6 \rangle$, where the required single-qubit operations leave the default measurement settings unchanged for Alice and the Bobs, but rotate Charlie and Debbie's measurement settings.
	To measure the observable $\langle \text{X}_1 \text{X}_2 \text{X}_5 \text{X}_6 \rangle$ of the GHZ state to obtain $Q_{\text{X}}$, all network users must measure the observable $\langle \text{X}_1 \text{Y}_2 \text{X}_3 \text{X}_4 \text{X}_5 \text{Y}_6 \rangle$, where now single-qubit operations encoded onto the measurement settings correspond to rotations of measurements into a different basis.
	When Alice and the Bobs wish to partake in QCKA with bi-partite resources, the measurement procedure suitably follows that of the GHZ state, except now the observables are for Bell states, and alternate single-qubit operations---based on required rotations applied to the initial graph---are encoded onto these settings.
	Full blueprints containing the operations applied to each qubit within each measurement set are presented in Figure~\ref{fig:expperimentalTechnique}.
	The measured $\langle \text{Z}^{\otimes \text{N}} \rangle$ and $\langle \text{X}^{\otimes \text{N}} \rangle$ observables, where $\text{N}$ is the qubit number, of the three Bell states and the GHZ state used in this protocol are presented in Figure~\ref{fig:results}.

    \vspace{1em}
    \noindent
    {\normalsize\textbf{Data availability}}
    \newline
    \noindent
    Data collected from the experimental work supporting the results reported in this manuscript, as well as the analysis, is available at \url{https://gitlab.com/EMQlab/conference-key-agreement-in-a-quantum-network.}
    }    


\bibliography{bib}

\begin{thebibliography}{10}
\expandafter\ifx\csname url\endcsname\relax
  \def\url#1{\texttt{#1}}\fi
\expandafter\ifx\csname urlprefix\endcsname\relax\def\urlprefix{URL }\fi
\providecommand{\bibinfo}[2]{#2}
\providecommand{\eprint}[2][]{\url{#2}}

\bibitem{beals_efficient_2013}
\bibinfo{author}{Beals, R.} \emph{et~al.}
\newblock \bibinfo{title}{Efficient distributed quantum computing}.
\newblock \emph{\bibinfo{journal}{Proc. R. Soc. A: Math. Phys. Eng. Sci.}}
  \textbf{\bibinfo{volume}{469}}, \bibinfo{pages}{20120686}
  (\bibinfo{year}{2013}).

\bibitem{van_meter_path_2016}
\bibinfo{author}{Van~Meter, R.} \& \bibinfo{author}{Devitt, S.~J.}
\newblock \bibinfo{title}{The path to scalable distributed quantum computing}.
\newblock \emph{\bibinfo{journal}{Computer}} \textbf{\bibinfo{volume}{49}},
  \bibinfo{pages}{31--42} (\bibinfo{year}{2016}).

\bibitem{gottesman2012}
\bibinfo{author}{Gottesman, D.}, \bibinfo{author}{Jennewein, T.} \&
  \bibinfo{author}{Croke, S.}
\newblock \bibinfo{title}{Longer-baseline telescopes using quantum repeaters}.
\newblock \emph{\bibinfo{journal}{Phys. Rev. Lett.}}
  \textbf{\bibinfo{volume}{109}}, \bibinfo{pages}{070503}
  (\bibinfo{year}{2012}).

\bibitem{komar_quantum_2014}
\bibinfo{author}{Kómár, P.} \emph{et~al.}
\newblock \bibinfo{title}{A quantum network of clocks}.
\newblock \emph{\bibinfo{journal}{Nat. Phys.}} \textbf{\bibinfo{volume}{10}},
  \bibinfo{pages}{582--587} (\bibinfo{year}{2014}).

\bibitem{bennett_quantum_2014}
\bibinfo{author}{Bennett, C.~H.} \& \bibinfo{author}{Brassard, G.}
\newblock \bibinfo{title}{Quantum cryptography: {Public} key distribution and
  coin tossing}.
\newblock \emph{\bibinfo{journal}{Theor. Comput. Sci.}}
  \textbf{\bibinfo{volume}{560}}, \bibinfo{pages}{7--11}
  (\bibinfo{year}{2014}).

\bibitem{ekert_quantum_1991}
\bibinfo{author}{Ekert, A.~K.}
\newblock \bibinfo{title}{Quantum cryptography based on {Bell}'s theorem}.
\newblock \emph{\bibinfo{journal}{Phys. Rev. Lett.}}
  \textbf{\bibinfo{volume}{67}}, \bibinfo{pages}{661--663}
  (\bibinfo{year}{1991}).

\bibitem{peev2009secoqc}
\bibinfo{author}{Peev, M.} \emph{et~al.}
\newblock \bibinfo{title}{The {SECOQC} quantum key distribution network in
  {Vienna}}.
\newblock \emph{\bibinfo{journal}{New. J. Phys.}}
  \textbf{\bibinfo{volume}{11}}, \bibinfo{pages}{075001}
  (\bibinfo{year}{2009}).

\bibitem{sasaki2011tokyo}
\bibinfo{author}{Sasaki, M.} \emph{et~al.}
\newblock \bibinfo{title}{Field test of quantum key distribution in the {Tokyo
  QKD} network}.
\newblock \emph{\bibinfo{journal}{Opt. Express.}}
  \textbf{\bibinfo{volume}{19}}, \bibinfo{pages}{10387--10409}
  (\bibinfo{year}{2011}).

\bibitem{dynes2019cambridge}
\bibinfo{author}{Dynes, J.} \emph{et~al.}
\newblock \bibinfo{title}{Cambridge quantum network}.
\newblock \emph{\bibinfo{journal}{npj Quantum Information}}
  \textbf{\bibinfo{volume}{5}}, \bibinfo{pages}{101} (\bibinfo{year}{2019}).

\bibitem{chen_implementation_2021}
\bibinfo{author}{Chen, T.-Y.} \emph{et~al.}
\newblock \bibinfo{title}{Implementation of a 46-node quantum metropolitan area
  network}.
\newblock \emph{\bibinfo{journal}{npj Quantum Information}}
  \textbf{\bibinfo{volume}{7}}, \bibinfo{pages}{134} (\bibinfo{year}{2021}).

\bibitem{epping_multi-partite_2017}
\bibinfo{author}{Epping, M.}, \bibinfo{author}{Kampermann, H.},
  \bibinfo{author}{Macchiavello, C.} \& \bibinfo{author}{Bruß, D.}
\newblock \bibinfo{title}{Multi-partite entanglement can speed up quantum key
  distribution in networks}.
\newblock \emph{\bibinfo{journal}{New. J. Phys.}}
  \textbf{\bibinfo{volume}{19}}, \bibinfo{pages}{093012}
  (\bibinfo{year}{2017}).

\bibitem{grasselli_finite-key_2018}
\bibinfo{author}{Grasselli, F.}, \bibinfo{author}{Kampermann, H.} \&
  \bibinfo{author}{Bruß, D.}
\newblock \bibinfo{title}{Finite-key effects in multipartite quantum key
  distribution protocols}.
\newblock \emph{\bibinfo{journal}{New. J. Phys.}}
  \textbf{\bibinfo{volume}{20}}, \bibinfo{pages}{113014}
  (\bibinfo{year}{2018}).

\bibitem{murta_quantum_2020}
\bibinfo{author}{Murta, G.}, \bibinfo{author}{Grasselli, F.},
  \bibinfo{author}{Kampermann, H.} \& \bibinfo{author}{Bruß, D.}
\newblock \bibinfo{title}{Quantum conference key agreement: a review}.
\newblock \emph{\bibinfo{journal}{Adv. Quantum Technol.}}
  \textbf{\bibinfo{volume}{3}}, \bibinfo{pages}{2000025}
  (\bibinfo{year}{2020}).

\bibitem{proietti_experimental_2021}
\bibinfo{author}{Proietti, M.} \emph{et~al.}
\newblock \bibinfo{title}{Experimental quantum conference key agreement}.
\newblock \emph{\bibinfo{journal}{Sci. Adv.}} \textbf{\bibinfo{volume}{7}},
  \bibinfo{pages}{eabe0395} (\bibinfo{year}{2021}).

\bibitem{hillery1999}
\bibinfo{author}{Hillery, M.}, \bibinfo{author}{Bu{\v{z}}ek, V.} \&
  \bibinfo{author}{Berthiaume, A.}
\newblock \bibinfo{title}{Quantum secret sharing}.
\newblock \emph{\bibinfo{journal}{Phys. Rev. A.}}
  \textbf{\bibinfo{volume}{59}}, \bibinfo{pages}{1829} (\bibinfo{year}{1999}).

\bibitem{tittel2001}
\bibinfo{author}{Tittel, W.}, \bibinfo{author}{Zbinden, H.} \&
  \bibinfo{author}{Gisin, N.}
\newblock \bibinfo{title}{Experimental demonstration of quantum secret
  sharing}.
\newblock \emph{\bibinfo{journal}{Phys. Rev. A.}}
  \textbf{\bibinfo{volume}{63}}, \bibinfo{pages}{042301}
  (\bibinfo{year}{2001}).

\bibitem{broadbent2009}
\bibinfo{author}{Broadbent, A.}, \bibinfo{author}{Chouha, P.-R.} \&
  \bibinfo{author}{Tapp, A.}
\newblock \bibinfo{title}{The {GHZ} state in secret sharing and entanglement
  simulation}.
\newblock In \emph{\bibinfo{booktitle}{2009 Third International Conference on
  Quantum, Nano and Micro Technologies}}, \bibinfo{pages}{59--62}
  (\bibinfo{organization}{IEEE}, \bibinfo{year}{2009}).

\bibitem{buhrman2010}
\bibinfo{author}{Buhrman, H.}, \bibinfo{author}{Cleve, R.},
  \bibinfo{author}{Massar, S.} \& \bibinfo{author}{De~Wolf, R.}
\newblock \bibinfo{title}{Nonlocality and communication complexity}.
\newblock \emph{\bibinfo{journal}{Rev. Mod. Phys.}}
  \textbf{\bibinfo{volume}{82}}, \bibinfo{pages}{665} (\bibinfo{year}{2010}).

\bibitem{ho2022}
\bibinfo{author}{Ho, J.} \emph{et~al.}
\newblock \bibinfo{title}{Entanglement-based quantum communication complexity
  beyond {Bell} nonlocality}.
\newblock \emph{\bibinfo{journal}{npj Quantum Information}}
  \textbf{\bibinfo{volume}{8}}, \bibinfo{pages}{13} (\bibinfo{year}{2022}).

\bibitem{wengerowsky_entanglement-based_2018}
\bibinfo{author}{Wengerowsky, S.}, \bibinfo{author}{Joshi, S.~K.},
  \bibinfo{author}{Steinlechner, F.}, \bibinfo{author}{Hübel, H.} \&
  \bibinfo{author}{Ursin, R.}
\newblock \bibinfo{title}{An entanglement-based wavelength-multiplexed quantum
  communication network}.
\newblock \emph{\bibinfo{journal}{Nature}} \textbf{\bibinfo{volume}{564}},
  \bibinfo{pages}{225--228} (\bibinfo{year}{2018}).

\bibitem{joshi_trusted_2020}
\bibinfo{author}{Joshi, S.~K.} \emph{et~al.}
\newblock \bibinfo{title}{A trusted node–free eight-user metropolitan quantum
  communication network}.
\newblock \emph{\bibinfo{journal}{Sci. Adv.}} \textbf{\bibinfo{volume}{6}},
  \bibinfo{pages}{eaba0959} (\bibinfo{year}{2020}).

\bibitem{hahn_limitations_2022-1}
\bibinfo{author}{Hahn, F.}, \bibinfo{author}{Dahlberg, A.},
  \bibinfo{author}{Eisert, J.} \& \bibinfo{author}{Pappa, A.}
\newblock \bibinfo{title}{Limitations of nearest-neighbor quantum networks}.
\newblock \emph{\bibinfo{journal}{Physical Review A}}
  \textbf{\bibinfo{volume}{106}}, \bibinfo{pages}{L010401}
  (\bibinfo{year}{2022}).

\bibitem{hein_multiparty_2004}
\bibinfo{author}{Hein, M.}, \bibinfo{author}{Eisert, J.} \&
  \bibinfo{author}{Briegel, H.~J.}
\newblock \bibinfo{title}{Multi-party entanglement in graph states}.
\newblock \emph{\bibinfo{journal}{Phys. Rev. A.}}
  \textbf{\bibinfo{volume}{69}}, \bibinfo{pages}{062311}
  (\bibinfo{year}{2004}).

\bibitem{hahn_quantum_2019}
\bibinfo{author}{Hahn, F.}, \bibinfo{author}{Pappa, A.} \&
  \bibinfo{author}{Eisert, J.}
\newblock \bibinfo{title}{Quantum network routing and local complementation}.
\newblock \emph{\bibinfo{journal}{npj Quantum Information}}
  \textbf{\bibinfo{volume}{5}}, \bibinfo{pages}{76} (\bibinfo{year}{2019}).

\bibitem{adcock_mapping_2020}
\bibinfo{author}{Adcock, J.~C.}, \bibinfo{author}{Morley-Short, S.},
  \bibinfo{author}{Dahlberg, A.} \& \bibinfo{author}{Silverstone, J.~W.}
\newblock \bibinfo{title}{Mapping graph state orbits under local
  complementation}.
\newblock \emph{\bibinfo{journal}{Quantum}} \textbf{\bibinfo{volume}{4}},
  \bibinfo{pages}{305} (\bibinfo{year}{2020}).

\bibitem{pirandola_direct_2009}
\bibinfo{author}{Pirandola, S.}, \bibinfo{author}{García-Patrón, R.},
  \bibinfo{author}{Braunstein, S.~L.} \& \bibinfo{author}{Lloyd, S.}
\newblock \bibinfo{title}{Direct and reverse secret-key capacities of a quantum
  channel}.
\newblock \emph{\bibinfo{journal}{Phys. Rev. Lett.}}
  \textbf{\bibinfo{volume}{102}}, \bibinfo{pages}{050503}
  (\bibinfo{year}{2009}).

\bibitem{pirandola_fundamental_2017}
\bibinfo{author}{Pirandola, S.}, \bibinfo{author}{Laurenza, R.},
  \bibinfo{author}{Ottaviani, C.} \& \bibinfo{author}{Banchi, L.}
\newblock \bibinfo{title}{Fundamental limits of repeaterless quantum
  communications}.
\newblock \emph{\bibinfo{journal}{Nat. Commun.}} \textbf{\bibinfo{volume}{8}},
  \bibinfo{pages}{15043} (\bibinfo{year}{2017}).

\bibitem{pirandola_end--end_2019-1}
\bibinfo{author}{Pirandola, S.}
\newblock \bibinfo{title}{End-to-end capacities of a quantum communication
  network}.
\newblock \emph{\bibinfo{journal}{Commun. Phys.}} \textbf{\bibinfo{volume}{2}},
  \bibinfo{pages}{1--10} (\bibinfo{year}{2019}).

\bibitem{pirandola_general_2020}
\bibinfo{author}{Pirandola, S.}
\newblock \bibinfo{title}{General upper bound for conferencing keys in
  arbitrary quantum networks}.
\newblock \emph{\bibinfo{journal}{IET Quantum Commun.}}
  \textbf{\bibinfo{volume}{1}}, \bibinfo{pages}{22--25} (\bibinfo{year}{2020}).

\bibitem{ottaviani_modular_2019-1}
\bibinfo{author}{Ottaviani, C.}, \bibinfo{author}{Lupo, C.},
  \bibinfo{author}{Laurenza, R.} \& \bibinfo{author}{Pirandola, S.}
\newblock \bibinfo{title}{Modular network for high-rate quantum conferencing}.
\newblock \emph{\bibinfo{journal}{Commun. Phys.}} \textbf{\bibinfo{volume}{2}},
  \bibinfo{pages}{1--6} (\bibinfo{year}{2019}).

\bibitem{miguel-ramiro_optimized_2023}
\bibinfo{author}{Miguel-Ramiro, J.}, \bibinfo{author}{Pirker, A.} \&
  \bibinfo{author}{Dür, W.}
\newblock \bibinfo{title}{Optimized quantum networks}.
\newblock \emph{\bibinfo{journal}{Quantum}} \textbf{\bibinfo{volume}{7}},
  \bibinfo{pages}{919} (\bibinfo{year}{2023}).

\bibitem{wallnofer_multipartite_2019}
\bibinfo{author}{Wallnöfer, J.}, \bibinfo{author}{Pirker, A.},
  \bibinfo{author}{Zwerger, M.} \& \bibinfo{author}{Dür, W.}
\newblock \bibinfo{title}{Multipartite state generation in quantum networks
  with optimal scaling}.
\newblock \emph{\bibinfo{journal}{Scientific Reports}}
  \textbf{\bibinfo{volume}{9}}, \bibinfo{pages}{314} (\bibinfo{year}{2019}).

\bibitem{kuzmin_scalable_2019}
\bibinfo{author}{Kuzmin, V.~V.}, \bibinfo{author}{Vasilyev, D.~V.},
  \bibinfo{author}{Sangouard, N.}, \bibinfo{author}{Dür, W.} \&
  \bibinfo{author}{Muschik, C.~A.}
\newblock \bibinfo{title}{Scalable repeater architectures for multi-party
  states}.
\newblock \emph{\bibinfo{journal}{npj Quantum Information}}
  \textbf{\bibinfo{volume}{5}}, \bibinfo{pages}{115} (\bibinfo{year}{2019}).

\bibitem{Fu_2015}
\bibinfo{author}{Fu, Y.}, \bibinfo{author}{Yin, H.-L.}, \bibinfo{author}{Chen,
  T.-Y.} \& \bibinfo{author}{Chen, Z.-B.}
\newblock \bibinfo{title}{Long-distance measurement-device-independent
  multiparty quantum communication}.
\newblock \emph{\bibinfo{journal}{Phys. Rev. Lett.}}
  \textbf{\bibinfo{volume}{114}}, \bibinfo{pages}{090501}
  (\bibinfo{year}{2015}).

\bibitem{grasselli_wStateCKA_2019}
\bibinfo{author}{Grasselli, F.}, \bibinfo{author}{Kampermann, H.} \&
  \bibinfo{author}{Bru{\ss}, D.}
\newblock \bibinfo{title}{Conference key agreement with single-photon
  interference}.
\newblock \emph{\bibinfo{journal}{New. J. Phys.}}
  \textbf{\bibinfo{volume}{21}}, \bibinfo{pages}{123002}
  (\bibinfo{year}{2019}).

\bibitem{Cao_2021}
\bibinfo{author}{Cao, X.-Y.}, \bibinfo{author}{Gu, J.}, \bibinfo{author}{Lu,
  Y.-S.}, \bibinfo{author}{Yin, H.-L.} \& \bibinfo{author}{Chen, Z.-B.}
\newblock \bibinfo{title}{Coherent one-way quantum conference key agreement
  based on twin field}.
\newblock \emph{\bibinfo{journal}{New. J. Phys.}}
  \textbf{\bibinfo{volume}{23}}, \bibinfo{pages}{043002}
  (\bibinfo{year}{2021}).

\bibitem{singkanipa2021}
\bibinfo{author}{Singkanipa, P.} \& \bibinfo{author}{Kok, P.}
\newblock \bibinfo{title}{Quantum conference key agreement with photon loss}.
\newblock \emph{\bibinfo{journal}{arXiv preprint:2101.01483}}
  (\bibinfo{year}{2021}).

\bibitem{carrara_genuine_2021}
\bibinfo{author}{Carrara, G.}, \bibinfo{author}{Kampermann, H.},
  \bibinfo{author}{Bruß, D.} \& \bibinfo{author}{Murta, G.}
\newblock \bibinfo{title}{Genuine multipartite entanglement is not a
  precondition for secure conference key agreement}.
\newblock \emph{\bibinfo{journal}{Phys. Rev. R.}} \textbf{\bibinfo{volume}{3}},
  \bibinfo{pages}{013264} (\bibinfo{year}{2021}).

\bibitem{Das_2021}
\bibinfo{author}{Das, S.}, \bibinfo{author}{B\"auml, S.},
  \bibinfo{author}{Winczewski, M.} \& \bibinfo{author}{Horodecki, K.}
\newblock \bibinfo{title}{Universal limitations on quantum key distribution
  over a network}.
\newblock \emph{\bibinfo{journal}{Phys. Rev. X}} \textbf{\bibinfo{volume}{11}},
  \bibinfo{pages}{041016} (\bibinfo{year}{2021}).

\bibitem{hahn2020}
\bibinfo{author}{Hahn, F.}, \bibinfo{author}{de~Jong, J.} \&
  \bibinfo{author}{Pappa, A.}
\newblock \bibinfo{title}{Anonymous quantum conference key agreement}.
\newblock \emph{\bibinfo{journal}{PRX Quantum}} \textbf{\bibinfo{volume}{1}},
  \bibinfo{pages}{020325} (\bibinfo{year}{2020}).

\bibitem{grasselli_secure_2022}
\bibinfo{author}{Grasselli, F.} \emph{et~al.}
\newblock \bibinfo{title}{Secure {{Anonymous Conferencing}} in {{Quantum
  Networks}}}.
\newblock \emph{\bibinfo{journal}{PRX Quantum}} \textbf{\bibinfo{volume}{3}},
  \bibinfo{pages}{040306} (\bibinfo{year}{2022}).

\bibitem{de2022anonymous}
\bibinfo{author}{de~Jong, J.}, \bibinfo{author}{Hahn, F.},
  \bibinfo{author}{Eisert, J.}, \bibinfo{author}{Walk, N.} \&
  \bibinfo{author}{Pappa, A.}
\newblock \bibinfo{title}{Anonymous conference key agreement in linear quantum
  networks}.
\newblock \emph{\bibinfo{journal}{arXiv preprint arXiv:2205.09169}}
  (\bibinfo{year}{2022}).

\bibitem{thalacker2021anonymous}
\bibinfo{author}{Thalacker, C.}, \bibinfo{author}{Hahn, F.},
  \bibinfo{author}{de~Jong, J.}, \bibinfo{author}{Pappa, A.} \&
  \bibinfo{author}{Barz, S.}
\newblock \bibinfo{title}{Anonymous and secret communication in quantum
  networks}.
\newblock \emph{\bibinfo{journal}{New. J. Phys.}}
  \textbf{\bibinfo{volume}{23}}, \bibinfo{pages}{083026}
  (\bibinfo{year}{2021}).

\bibitem{huang2022_anonymous}
\bibinfo{author}{Huang, Z.} \emph{et~al.}
\newblock \bibinfo{title}{Experimental implementation of secure anonymous
  protocols on an eight-user quantum key distribution network}.
\newblock \emph{\bibinfo{journal}{npj Quantum Information}}
  \textbf{\bibinfo{volume}{8}}, \bibinfo{pages}{25} (\bibinfo{year}{2022}).

\bibitem{walk2021}
\bibinfo{author}{Walk, N.} \& \bibinfo{author}{Eisert, J.}
\newblock \bibinfo{title}{Sharing classical secrets with continuous-variable
  entanglement: Composable security and network coding advantage}.
\newblock \emph{\bibinfo{journal}{PRX Quantum}} \textbf{\bibinfo{volume}{2}},
  \bibinfo{pages}{040339} (\bibinfo{year}{2021}).

\bibitem{liu_distributed_2021}
\bibinfo{author}{Liu, L.-Z.} \emph{et~al.}
\newblock \bibinfo{title}{Distributed quantum phase estimation with entangled
  photons}.
\newblock \emph{\bibinfo{journal}{Nat. Photonics.}}
  \textbf{\bibinfo{volume}{15}}, \bibinfo{pages}{137--142}
  (\bibinfo{year}{2021}).

\bibitem{dahlberg_how_2020}
\bibinfo{author}{Dahlberg, A.}, \bibinfo{author}{Helsen, J.} \&
  \bibinfo{author}{Wehner, S.}
\newblock \bibinfo{title}{How to transform graph states using single-qubit
  operations: computational complexity and algorithms}.
\newblock \emph{\bibinfo{journal}{Quantum Sci. Technol.}}
  \textbf{\bibinfo{volume}{5}}, \bibinfo{pages}{045016} (\bibinfo{year}{2020}).

\bibitem{pickston_optimised_2021}
\bibinfo{author}{Pickston, A.} \emph{et~al.}
\newblock \bibinfo{title}{Optimised domain-engineered crystals for pure telecom
  photon sources}.
\newblock \emph{\bibinfo{journal}{Opt. Express.}}
  \textbf{\bibinfo{volume}{29}}, \bibinfo{pages}{6991--7002}
  (\bibinfo{year}{2021}).

\bibitem{fedrizzi_wavelength-tunable_2007}
\bibinfo{author}{Fedrizzi, A.}, \bibinfo{author}{Herbst, T.},
  \bibinfo{author}{Poppe, A.}, \bibinfo{author}{Jennewein, T.} \&
  \bibinfo{author}{Zeilinger, A.}
\newblock \bibinfo{title}{A wavelength-tunable fiber-coupled source of
  narrowband entangled photons}.
\newblock \emph{\bibinfo{journal}{Opt. Express.}}
  \textbf{\bibinfo{volume}{15}}, \bibinfo{pages}{15377--15386}
  (\bibinfo{year}{2007}).

\bibitem{browne_resource-efficient_2005}
\bibinfo{author}{Browne, D.~E.} \& \bibinfo{author}{Rudolph, T.}
\newblock \bibinfo{title}{Resource-efficient linear optical quantum
  computation}.
\newblock \emph{\bibinfo{journal}{Phys. Rev. Lett.}}
  \textbf{\bibinfo{volume}{95}}, \bibinfo{pages}{010501}
  (\bibinfo{year}{2005}).

\end{thebibliography}


\begin{thebibliography}{1}
\expandafter\ifx\csname url\endcsname\relax
  \def\url#1{\texttt{#1}}\fi
\expandafter\ifx\csname urlprefix\endcsname\relax\def\urlprefix{URL }\fi
\providecommand{\bibinfo}[2]{#2}
\providecommand{\eprint}[2][]{\url{#2}}

\bibitem{ralph_linear_2002-1}
\bibinfo{author}{Ralph, T.~C.}, \bibinfo{author}{Langford, N.~K.},
  \bibinfo{author}{Bell, T.~B.} \& \bibinfo{author}{White, A.~G.}
\newblock \bibinfo{title}{Linear optical controlled-{NOT} gate in the
  coincidence basis}.
\newblock \emph{\bibinfo{journal}{Phys. Rev. A.}}
  \textbf{\bibinfo{volume}{65}}, \bibinfo{pages}{062324}
  (\bibinfo{year}{2002}).

\bibitem{hofmann_quantum_2002}
\bibinfo{author}{Hofmann, H.~F.} \& \bibinfo{author}{Takeuchi, S.}
\newblock \bibinfo{title}{Quantum phase gate for photonic qubits using only
  beam splitters and postselection}.
\newblock \emph{\bibinfo{journal}{Phys. Rev. A.}}
  \textbf{\bibinfo{volume}{66}}, \bibinfo{pages}{024308}
  (\bibinfo{year}{2002}).

\bibitem{langford_demonstration_2005}
\bibinfo{author}{Langford, N.~K.} \emph{et~al.}
\newblock \bibinfo{title}{Demonstration of a simple entangling optical gate and
  its use in {Bell}-state analysis}.
\newblock \emph{\bibinfo{journal}{Phys. Rev. Lett.}}
  \textbf{\bibinfo{volume}{95}}, \bibinfo{pages}{210504}
  (\bibinfo{year}{2005}).

\bibitem{pickston_optimised_2021}
\bibinfo{author}{Pickston, A.} \emph{et~al.}
\newblock \bibinfo{title}{Optimised domain-engineered crystals for pure telecom
  photon sources}.
\newblock \emph{\bibinfo{journal}{Opt. Express.}}
  \textbf{\bibinfo{volume}{29}}, \bibinfo{pages}{6991--7002}
  (\bibinfo{year}{2021}).

\bibitem{fedrizzi_wavelength-tunable_2007}
\bibinfo{author}{Fedrizzi, A.}, \bibinfo{author}{Herbst, T.},
  \bibinfo{author}{Poppe, A.}, \bibinfo{author}{Jennewein, T.} \&
  \bibinfo{author}{Zeilinger, A.}
\newblock \bibinfo{title}{A wavelength-tunable fiber-coupled source of
  narrowband entangled photons}.
\newblock \emph{\bibinfo{journal}{Opt. Express.}}
  \textbf{\bibinfo{volume}{15}}, \bibinfo{pages}{15377--15386}
  (\bibinfo{year}{2007}).

\bibitem{dahlberg_transforming_2020}
\bibinfo{author}{Dahlberg, A.}, \bibinfo{author}{Helsen, J.} \&
  \bibinfo{author}{Wehner, S.}
\newblock \bibinfo{title}{Transforming graph states to {Bell}-pairs is
  {NP}-{Complete}}.
\newblock \emph{\bibinfo{journal}{Quantum}} \textbf{\bibinfo{volume}{4}},
  \bibinfo{pages}{348} (\bibinfo{year}{2020}).

\bibitem{dahlberg_how_2018}
\bibinfo{author}{Dahlberg, A.}, \bibinfo{author}{Helsen, J.} \&
  \bibinfo{author}{Wehner, S.}
\newblock \bibinfo{title}{How to transform graph states using single-qubit
  operations: computational complexity and algorithms}.
\newblock \emph{\bibinfo{journal}{arXiv:1805.05306 [quant-ph]}}
  (\bibinfo{year}{2018}).

\bibitem{dahlberg_complexity_2019}
\bibinfo{author}{Dahlberg, A.}, \bibinfo{author}{Helsen, J.} \&
  \bibinfo{author}{Wehner, S.}
\newblock \bibinfo{title}{The complexity of the vertex-minor problem}.
\newblock \emph{\bibinfo{journal}{arXiv:1906.05689 [quant-ph]}}
  (\bibinfo{year}{2019}).

\end{thebibliography}
\bibliographystyle{naturemag}

\newpage
\clearpage
\newpage

\end{document}


\title{\normalfont\LARGE{
Supplementary Information for ``Conference key agreement in a quantum network''
}}

\author{Alexander Pickston$^{\dagger}$}
\affiliation{Institute of Photonics and Quantum Sciences, School of Engineering and Physical Sciences, Heriot-Watt University, Edinburgh EH14 4AS, United Kingdom}

\author{Joseph Ho$^{\dagger}$}
\affiliation{Institute of Photonics and Quantum Sciences, School of Engineering and Physical Sciences, Heriot-Watt University, Edinburgh EH14 4AS, United Kingdom}

\author{Andrés Ulibarrena}
\affiliation{Institute of Photonics and Quantum Sciences, School of Engineering and Physical Sciences, Heriot-Watt University, Edinburgh EH14 4AS, United Kingdom}

\author{Federico Grasselli}
\affiliation{Institut für Theoretische Physik III, Heinrich-Heine-Universität Düsseldorf, Universitätsstraße 1, D-40225 Düsseldorf, Germany}

\author{Massimiliano Proietti}
\affiliation{Institute of Photonics and Quantum Sciences, School of Engineering and Physical Sciences, Heriot-Watt University, Edinburgh EH14 4AS, United Kingdom}

\author{Christopher L. Morrison}
\affiliation{Institute of Photonics and Quantum Sciences, School of Engineering and Physical Sciences, Heriot-Watt University, Edinburgh EH14 4AS, United Kingdom}

\author{Peter Barrow}
\affiliation{Institute of Photonics and Quantum Sciences, School of Engineering and Physical Sciences, Heriot-Watt University, Edinburgh EH14 4AS, United Kingdom}

\author{Francesco Graffitti}
\affiliation{Institute of Photonics and Quantum Sciences, School of Engineering and Physical Sciences, Heriot-Watt University, Edinburgh EH14 4AS, United Kingdom}

\author{Alessandro Fedrizzi}
\affiliation{Institute of Photonics and Quantum Sciences, School of Engineering and Physical Sciences, Heriot-Watt University, Edinburgh EH14 4AS, United Kingdom}
    
\maketitle

\subsection{\textbf{Supplementary Methods}}
    Here we outline how the six-qubit graph state in the experiment is mapped from the graph notation onto the linear optical circuit shown in Supplementary Figure~\ref{fig:SM_graph_circuit_equiv}.
    From the mathematical description of a graph (Equation~6 in the Methods of the main text) one can see that to prepare a graph state requires $V$ qubits initialised as $\ket{+}$ (equivalent to $\ket{\textsf{d}}$ in polarisation encoding), connected via \textsf{CZ} gates between specific qubits classified by $E$.
    The graph state generated in this work can be mapped onto a general circuit model depiction which consists of six qubits initialised in $|+\rangle$ and five \textsf{CZ} gates.
    Directly translating the circuit model depiction into an experimentally realisable linear optical circuit leads to large inefficiencies. 
    Realising \textsf{CZ} gates with linear optics results in a low probability of success of 1/9 per gate~\cite{ralph_linear_2002-1,hofmann_quantum_2002,langford_demonstration_2005}.
    Instead, we found a linear optical circuit that could produce a state equivalent to the targeted graph state (up to local rotations) based on linear optical fusion gates, which have a much higher probability of success of 1/2.
    Explicitly realising this linear optical circuit does require some entangled resources, namely two Bell pairs as well as a pair of separable photons.
    Three linear optical fusion gates are then used to generate multi-photon entanglement, producing the desired state.
    Finding the translation to a linear optical circuit enabled the investigation in this work, as the general circuit model approach carries a probability of success of 1/59,049.
    Our arrangement has a probability of success of 1/8.
    
    \begin{figure}[b!]
        \centering
        \includegraphics[width=0.5\columnwidth]{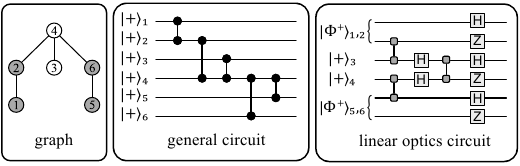}
        \caption{\textbf{Graph representation and its realisation.} 
        The 6 qubit graph state in our experiment is shown in the left panel. This can be translated into a generalised quantum circuit depicted in the middle panel. Our setup, which is more experimentally friendly in linear optics, is shown in the right panel. Notably we employ Bell states, $|\Phi^+\rangle$ and replace the two-qubit \textsf{CZ} gates with fusion gates. Additional single qubit operations, Hadamard and Pauli-Z gates are needed to correct the final state.
        }
        \label{fig:SM_graph_circuit_equiv}
    \end{figure}  
    
    We generate photon pairs using type-II parametric down-conversion (PDC) sources that operate in the filter-less regime, owed to the fact we employ domain engineering techniques to modify the phase-matching function (PMF) of the down-conversion process. 
    Despite our aperiodically poled KTP (aKTP) crystals operating without filters, two-photon interference visibilities of up to $98.6\pm1.1\%$ are still obtainable, meaning we carry forward minimal errors after any instances of two-photon interference.
    Without lossy spectral filters we can also access heralding efficiencies of up to $67.5\%$~\cite{pickston_optimised_2021}.
    
    We use a pulsed Ti:saphh laser with a central wavelength of 774.9nm producing photon pairs in a degenerate condition at 1549.8nm.
    These wavelengths were chosen to keep the crystal temperatures sufficiently far away from room temperature.
    This laser operates with a pulse duration of 1.3ps, matching the pump-envelope function (PEF) bandwidth to the PMF bandwidth at a repetition rate of 80MHz.
    
    For generating Bell states, we place our PDC sources inside a Sagnac interferometer, where a pump photon generates signal and idler pairs in either a clockwise or anti-clockwise direction depending on the polarisation state of the pump photon.
    When the pump photon is equally likely to produce photons in the clockwise direction and the anti-clockwise direction an entangled state is generated.
    This configuration explicitly consists of a Glan-Taylor prism, which ensures the pump is in a linear polarisation state followed by a HWP which rotates this linear polarisation state into $\ket{\textsf{d}}$ for generating an entangled state, or a linear basis state for generating a separable state.
    After the pump polarisation state is prepared, it is transmitted through a dichroic mirror (DM) which only reflects the down converted photons propagating back towards the pump.
    The pump is then probabilistically split on a dual-wavelength polarising beam-splitter (dPBS) into transmitted or reflected spatial modes.
    In the transmitted mode, the pump photon is in the $\ket{\textsf{h}_p}$ polarisation state and will generate signal and idler pairs in a clockwise direction.
    In the reflected mode, the pump photon is rotated from $\ket{\textsf{v}_p}$ into $\ket{\textsf{h}_p}$ by a dual-wavelength HWP (dHWP) to facilitate type-II PDC in the anti-clockwise direction.
    When photons are generated from a pump photon in the transmitted mode of the dPBS, the down-converted pair are each rotated by the dHWP, which acts to fix the temporal offset the different set of photon pairs possess, before being split by the dPBS and collected into SMFs~\cite{fedrizzi_wavelength-tunable_2007}.
    Upon collection, the signal and idler pairs exist in a maximally entangled state
    $$
    \ket{\Psi^-} = \frac{1}{\sqrt{2}} \ket{\textsf{h}_s\textsf{v}_i} - e^{i\phi} \ket{\textsf{v}_s\textsf{h}_i}.
    $$
    \begin{figure*}[t!]
        \centering
        \includegraphics{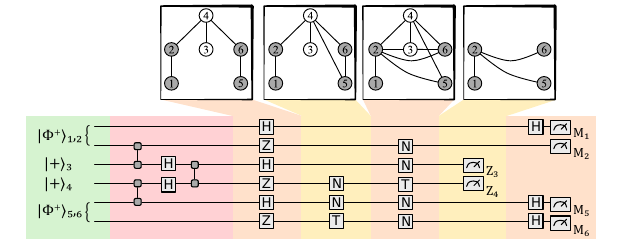}
        \caption{\textbf{Generating and manipulating the 6-photon graph state.} 
        We prepare the 6-photon graph state using photon-pair sources and a linear optics scheme. Two of the sources are prepared in the Bell state, $|\Phi^+\rangle$, while the third is in a biseparable state with each photon in $|+\rangle$ as shown in the green region. Using three Fusion gates and single-qubit rotations one arrives at the 6-photon graph state, up to a set of local rotations, as shown in the red region. The set of graph transformations for obtaining the 4-qubit GHZ state, in modes $\{1,2,5,6\}$, is shown in the two-tone orange shaded regions. Notably these consist of local operations and quantum measurements on non-participatory qubits $\{3,4\}$ in the Z basis.
        }
        \label{fig:SM_graph_scheme}
    \end{figure*}
    
    To generate the six-photon state presented in this work, we prepare two Bell states as $\ket{\Phi^+}$ (which are equivalent to $\ket{\Psi^-}$ up to local rotations) in modes $\{1,2\}$ and $\{5,6\}$ and a single separable state prepared as $\ket{\textsf{dd}}$ in modes $\{3,4\}$.
    We employ linear plate polarisers (LPPs) to prepare the state $\ket{\textsf{dd}}$.
    The required unitaries for Bell state preparation are performed with manual fibre polarisation controllers (FPCs).
    These unitaries act between the sources and fusion gates.
    The fusion gates consist of imposing two input optical modes, launched from SMFs, onto a PBS.
    Each output modes of the PBS are then collected into SMFs.
    A successful fusion operation occurs when there is one and only one photon in each output optical mode.
    This means that for the full circuit to be successful, only a single photon should be in the six optical modes after the three fusion gates. 
    Within each fusion gate there are additional optics to correct for the local phase that the PBS within each of these gates introduces.
    The correction optics are made up of a HWP sandwiched between two QWPs.
    Fusion gate 1 ($F1$) acts between modes $\{2,3\}$, fusion gate 2 ($F2$) between $\{4,5\}$ and finally, fusion gate 3 ($F3$) acts between one of the outputs of both $F1$ and $F2$, in modes $\{3,4\}$.
    The only difference between these gates is the presence of an additional HWP, just before the PBS within $F3$, which is present to apply a single-qubit Hadamard gate.
    Final FPCs situated directly after the fusion gates ensure that photons undergo an identity operation between the fusion gates and measurement.
    Only the detection of a six-fold event (six photons arriving individually across all six optical modes comprised of daughter PDC photons created from the same pump pulse train) heralds the success of all three fusion gates, even in our case where the outputs of $F1$ and $F2$ feed directly into the input of $F3$.
    
    The final part of the circuit is comprised of measurement stages which consist of a QWP, HWP and PBS. 
    These stages perform projective measurements. 
    The measured photons are detected using superconducting nano-wire single photon detectors that operate at quantum efficiencies $>80\%$ at our wavelengths.
    Detection events are processed by time-tagging logic. 
    A full schematic of the experimental arrangement can be seen in Supplementary Figure~\ref{fig:SM_expLayout}.
    \begin{figure*}[ht!]
        \centering
        \includegraphics{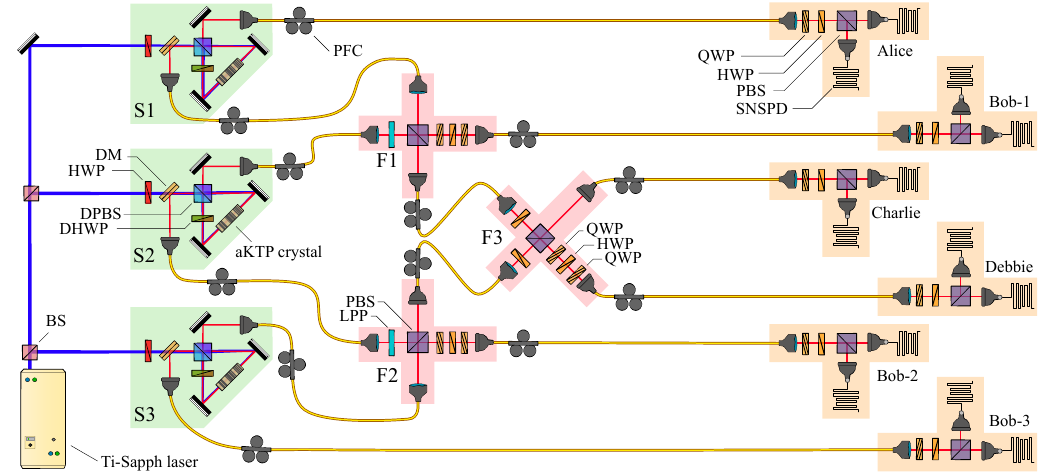}
        \caption{\textbf{Experimental layout.} 
        Highlighted in green are the sources, linear optical Fusion gates are in red and polarisation projective measurement stages in orange. Each source consists of an aKTP crystal, pumped by a Ti-Sapph laser and embedded in a Sagnac loop with a HWP to set the pump direction. Note only sources $\{S1,S3\}$ are entangled while $S2$ is pumped in one direction to produce biseparable states. Each photon is coupled into single-mode fibre with fibre polarisation controllers (PFC) used to set the desired polarisation states entering the Fusion gates $\{F1,F2\}$. Additionally, photons from $S2$ are sent through linear plate polarisers (LPP) to initialise their states in $|+\rangle$. Each Fusion gate comprises of two input optical modes which overlap on a polarising beam splitter (PBS) and successful operation of the two-qubit gate is conditional on one photon in each output mode. In each Fusion gate a HWP sandwiched by two quarter-wave plates (QWP) is used to compensate local phase shifts introduced by the PBS. In the final Fusion gate $F3$, two HWPs are used to apply the single-qubit Hadamard gates. The remaining single-qubit gates are determined by the protocol and implemented in the projective measurement stages on each qubit which consist of a QWP, HWP and PBS. Single photons are detected using superconducting nanowire single photon detectors (SNSPDs).
        }
        \label{fig:SM_expLayout}
    \end{figure*}
	
	All rotations required to allow Alice and Bobs to share different resource states among them are encoded onto the measurement settings. 
	Supplementary Figure~\ref{fig:SM_graph_scheme} is an example of the experimental arrangement for a round in which Alice and the Bobs wish to share a GHZ state.
	The single-qubit operations acting on each mode after the creation of the resource state, highlighted in yellow and orange within Supplementary Figure~\ref{fig:SM_graph_scheme}, are applied to each qubit upon measurement, meaning that each user is provided with measurement settings required to extract different states from the shared network resource state. 
	In this way, entanglement amongst a subset of the six network users can be allocated in a dynamic way.
	To determine the correct measurements users have to perform, we first decide which type of measurement round being performed, either type-1 or type-2.
	For both measurement types we already know that Charlie and Debbie, modes \{3,4\}, are required to measure in the Z-basis, so operations on their qubits are encoded onto their measurement basis.
	Settings on Alice and Bobs qubits, or modes \{1,2,5,6\}, do depend on which type of round is being performed. 
	If a type-1 round is taking place, then their starting measurement basis is Z, and the single qubit operations on that mode are applied to rotate (or leave unchanged) this measurement basis.
	Likewise for type-2 rounds.
	Hence, as shown in the Methods section, the measurement $\langle \text{Z}_1 \text{Z}_2 \text{Z}_3 \text{Z}_4 \text{Z}_5 \text{Z}_6 \rangle$ to complete a type-1 round actually requires the measurement $\langle \text{Z}_1 \text{Z}_2 \text{X}_3 \text{X}_4 \text{Z}_5 \text{Z}_6 \rangle$ after rotation have been applied, and the measurement $\langle \text{X}_1 \text{X}_2 \text{Z}_3 \text{Z}_4 \text{X}_5 \text{X}_6 \rangle$ becomes $\langle \text{X}_1 \text{Y}_2 \text{X}_3 \text{X}_4 \text{X}_5 \text{Y}_6 \rangle$.
    As a consequence of the GME implicit of the shared network resource state, whenever Charlie and Debbie perform a measurement they need to disentangle themselves from the system coherently. 
	When Charlie and Debbie measurements are in the $\text{X}$ basis (due to rotations), they can attain either 0 or 1 as their result. 
	The resulting GHZ may contain a local phase rotation conditioned on the outcomes of Charlie and Debbie. 
	This rotation must be corrected for and is done so in a post-processing step.
    The measurement results used for constructing and then measuring the correct noise parameters for each of the distilled states are shown in Supplementary Figure~\ref{fig:SM_allProjectionss}
    \begin{sidewaysfigure}[ht!]
       \centering
       \includegraphics[width=\textwidth]{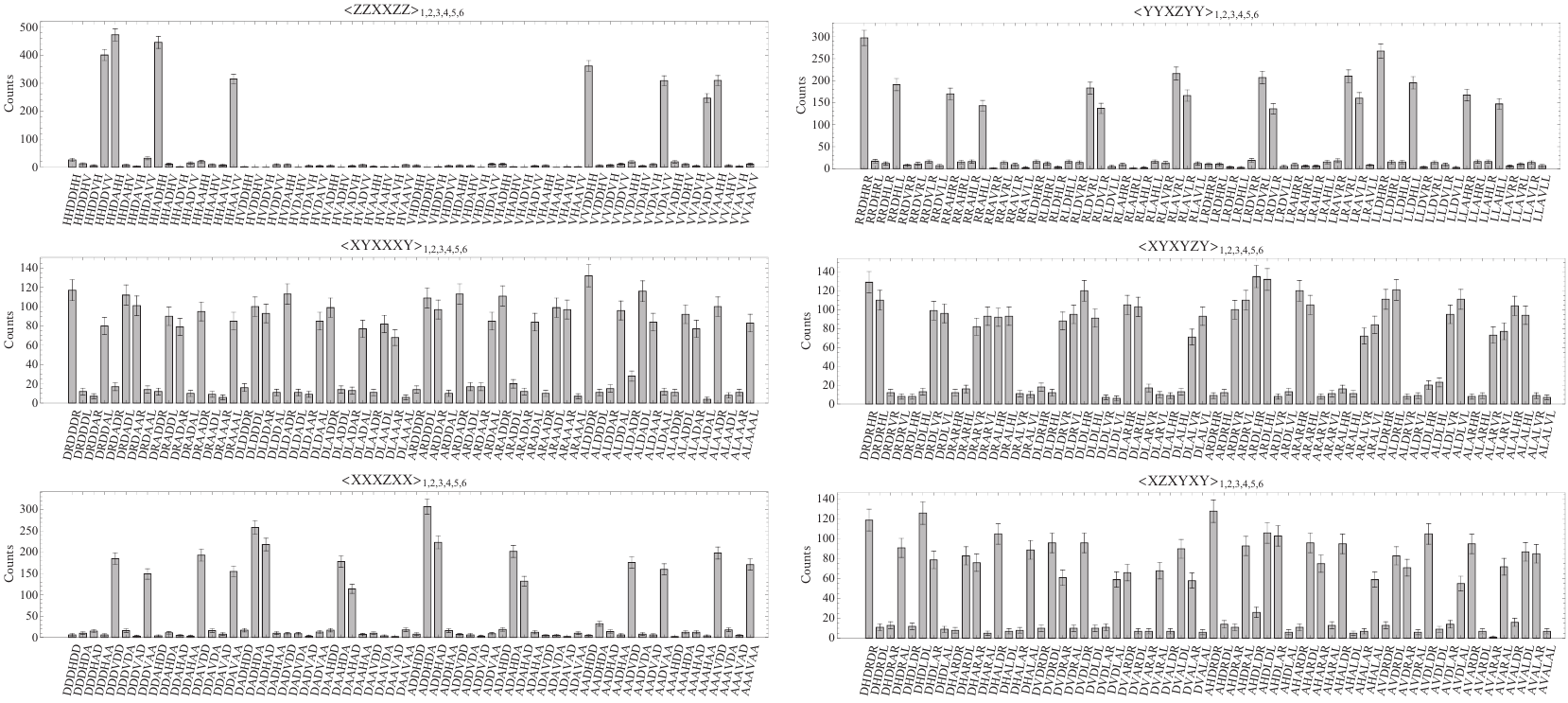}
       \caption{\textbf{Measured counts from projections.} 
        To obtain key rates from states derived from the network resource, specific observables need to be measured.  
        Contained in the plots are the 6-qubit projections expressed in their polarisation bases. 
        Error in count rates are calculated based on Poissonian statistics.}
        \label{fig:SM_allProjectionss}
    \end{sidewaysfigure}
	
	The work contained within this manuscript focuses on an investigation into a network protocol, where the particular scenario demands the distribution of a GHZ and a number of Bell states to specific users. 
	For the determining how to obtain states, we use a software library which can manipulate graph states based on available experimental transformations.
	This lets us form a search of available smaller resource states a given graph can produce.
    The search is brute-force, emanating from the fact that transforming graphs, or more generally stabilizer states, into tensor products of bipartite Bell pairs, or just a set of Bell pairs between specific vertices, using only a certain class of local operations and classical communication is a $\mathds{NP}$-complete problem~\cite{dahlberg_transforming_2020}.
	Likewise for GHZ state distillation from arbitrary graphs, if we wish to arrive at a GHZ state between a specific set of vertices~\cite{dahlberg_how_2018} or if we require a GHZ state of fixed size~\cite{dahlberg_complexity_2019}, the task is $\mathds{NP}$-complete.

\begin{figure*}[h!]
        \centering
        \includegraphics[width=\textwidth]{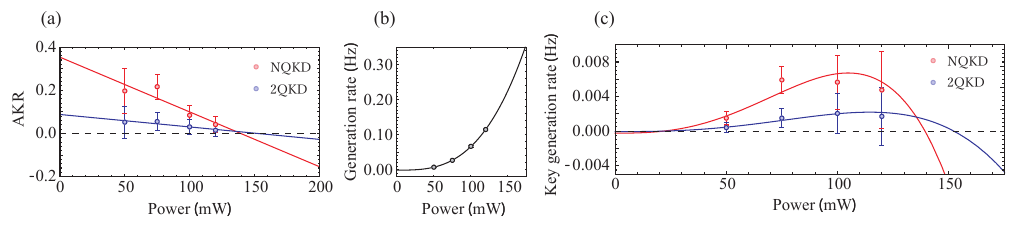}
        \caption{\textbf{AKR, raw photon generation rate and key generation rate as a function of the pump power.}
        (a) Data points correspond to the AKR from experimentally measured noise terms for NQKD and 2QKD. For each dataset a linear fit has been used to extrapolate the expected performance.
        (b) The measured 6-fold rate as a function of pump power is shown. A line of best fit assuming a cubic function has been used.
        (c) The key generation rate is a product of both the AKR and the raw generation rate. The ideal pump power is $\sim105$~mW in order to obtain the highest overall key generation rate for QCKA.}
        \label{fig:supp_powerSeries}
    \end{figure*}

    Assessing the performance of the protocol at a variety of different six-photon generation rates, reveals how the rates scale with noise.
    The main source of noise that degrades the quality of the prepared state is from higher order photon pair emissions due to the probabilistic nature of PDC sources.
    Even whilst operating with a post-selection criteria, higher order emission events can trigger detection patterns identical to that of the signal.
    These higher order emission terms result in a reduction to the magnitude of coherence, as events from incorrect pair generation events interfere with signal photons and degrade the purity of the state.
    The relation between the overall noise and the AKR is not trivial, partly due to the fact noise is not symmetrically distributed across the quantum state.
    This is evidenced by the different QBER and $Q_{\text{X}}$ for each of the Bell states derived from the same resource state.
    Multi-photon generation rate from a PDC source is not linear with respect to the amount of optical pump power applied to the non-linear crystal.
    The benefit of having higher pump powers is a larger $\text{N}$-fold generation rate.
    A caveat to this relationship is the additional noise generated by the presence of $> \text{N}$ photons consequential of the probabilistic nature of the PDC sources.
    As such, to attain the highest possible key generation rate, a power dependence measurement is required---this will reveal the ideal power to operate the protocol at.
    By computing the product of the AKR, which will fall with respect to an increase in pump power, and the raw photon generation rate, which will increase as a function of the pump power, the key generation rate can be calculated.
    Supplementary Figure~\ref{fig:supp_powerSeries} presents the scaling of AKR, generation rate and ultimately the key generation rate as a function of pump power, revealing that the highest key generation rate is at $\sim$105~mW.
    
\subsection{\textbf{Supplementary Discussion}}
    We remark that owing to the non-deterministic operation of the fusion gates and photon sources in this experiment, all six users must detect a single photon in their measurement stages to obtain the correct graph state.
    It would be an interesting prospect to determine whether this graph state inherits any loss robustness for the purposes of QCKA, or indeed other applications when deterministic sources and gates are used to generate them.
    Starting form $\ket{\textsf{G}}_{\text{initial}}$, to achieve two Bell states between $\{1,2\}$ and $\{5,6\}$ all that is required---in principle---is a single $\text{Z}$ basis measurement on qubit 4, tolerating the loss of qubit 3 and still adhering to the graph state formalism.
    \begin{figure}[h!]
        \centering
        \includegraphics[width=0.35\columnwidth]{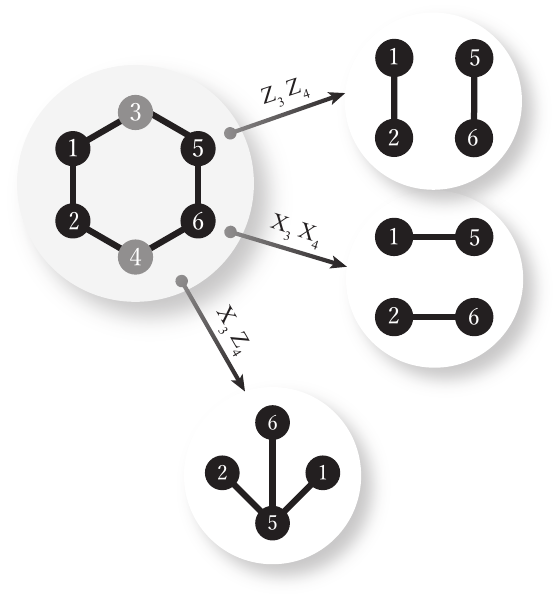}
        \caption{\textbf{Ring graph state and operations required to obtain smaller network resource states.}\label{fig:supp_ring_to_resources}
        The ring graph can be used in a similar means to $\ket{\textsf{G}}_{\text{initial}}$ to compare the key rates of different approaches to QCKA.
        Four users---in this illustration \{1,~3,~5,~6\}---are designated as the network users.
        These four users can attain pair-wise correlations with each other upon the application of Z measurements or X measurements by the two remaining users.
        Whilst there are several ways to obtain a GHZ state from the ring graph, we show how to obtain the state from measurements in the Pauli basis.}
    \end{figure}

    In this work, we chose to generate the resource graph state $\ket{\textsf{G}}_{\text{initial}}$ based on its accessibility with probabilistic photon sources and linear optical components.
    If we had generated a $\text{N}$ qubit GHZ state and distributed this amongst network users, the rate advantage for NQKD scales as $\text{N}-1$.
    We however, showed a rate advantage for the NQKD approach which did not scale as $\text{N}-1$, but a rate advantage of a factor of two, due to the ability to multi-cast Bell pairs from the shared resource state.
    Interestingly, there are families of graph states where the rate advantage can be reduced even further.
    In a similar scenario to the work in the main text, the ring graph is capable of distributing a GHZ state and Bell states to network users.
    Using the ring graph as a network resources can allow users to cast pairs of Bell states in \textit{every} round, differing from the graph used for investigations within this work.
    Because of this, there is more contribution to the overall key per network usage due to more pair-wise key distribution within the 2QKD rounds.
    Supplementary Figure~\ref{fig:supp_ring_to_resources} shows the operations required to obtain smaller network resources from the ring graph state. 
    Like in the main text, all operations are either local rotations or single-qubit projection measurements.

\bibliography{bib}
\bibliographystyle{naturemag}